\documentclass[twocolumn,english,aps,pra,showpacs]{revtex4}
\usepackage[T1]{fontenc}
\usepackage[latin1]{inputenc}
\usepackage{amsmath}
\usepackage{amssymb}
\usepackage{graphicx}
\usepackage{esint}

\makeatletter
\@ifundefined{textcolor}{}
{%
 \definecolor{BLACK}{gray}{0}
 \definecolor{WHITE}{gray}{1}
 \definecolor{RED}{rgb}{1,0,0}
 \definecolor{GREEN}{rgb}{0,1,0}
 \definecolor{BLUE}{rgb}{0,0,1}
 \definecolor{CYAN}{cmyk}{1,0,0,0}
 \definecolor{MAGENTA}{cmyk}{0,1,0,0}
 \definecolor{YELLOW}{cmyk}{0,0,1,0}
}

\makeatother

\makeatother

\usepackage{babel}
\begin{document}

\title{Critical temperature of a Rashba spin-orbit coupled Bose gas in harmonic
traps}

\author{Hui Hu$^{1}$ and Xia-Ji Liu$^{1}$ }

\affiliation{$^{1}$ACQAO and Centre for Atom Optics and Ultrafast Spectroscopy,
Swinburne University of Technology, Melbourne 3122, Australia}

\date{\today}
\begin{abstract}
We investigate theoretically Bose-Einstein condensation of an ideal,
trapped Bose gas in the presence of Rashba spin-orbit coupling. Analytic
results for the critical temperature and condensate fraction are derived,
based on a semi-classical approximation to the single-particle energy
spectrum and density of states, and are compared with exact results
obtained by explicitly summing discrete energy levels for small number
of particles. We find a significant decrease of the critical temperature
and of the condensate fraction due to a finite spin-orbit coupling.
For large coupling strength and finite number of particles $N$, the
critical temperature scales as $N^{2/5}$ and $N^{2/3}$ in three
and two dimensions, respectively, contrasted to the predictions of
$N^{1/3}$ and $N^{1/2}$ in the absence of spin-orbit coupling. Finite
size corrections in three dimensions are also discussed. 
\end{abstract}


\maketitle

\section{Introduction}

The recent experiment on spin-orbit (SO) coupled spinor Bose gases
of $^{87}$Rb atoms \cite{Lin2011} has stimulated great interest
in the theoretical study of SO physics in both Bose-Einstein condensation
(BEC) and fermionic superfluidity. It is well-known that the SO coupling
leads to many interesting phenomena in condensed matter physics. A
typical example is the recently discovered topological insulators
or quantum spin Hall states \cite{Qi2010,Hasan2010}. In degenerate
atomic gases, due to unprecedented controllability in the interatomic
interaction, geometry and purity \cite{Dalfovo1999,Bloch2008}, the
SO coupling may bring even more intriguing states of matter \cite{Zhai2011,Stanescu2008,Larson2009,Wang2010,Wu2011,Ho2011,Xu2011,Hu2011a,Sinha2011,Barnett2011,Zhu2011a,Deng2011,Vyasanakere2011,Iskin2011,Zhu2011b,Yu2011,Hu2011b,Gong2011,Liu2011}.

For a SO coupled BEC, non-trivial structures, such as the density-stripe
state \cite{Wang2010,Ho2011}, half-quantum vortex state \cite{Wu2011}
and lattice state \cite{Hu2011a,Sinha2011}, are predicted. For an
atomic Fermi gas near Feshbach resonances, new two-fermion bound states
with anisotropic mass are formed even at a negative $s$-wave scattering
length \cite{Vyasanakere2011,Yu2011,Hu2011b}, leading to the prospect
of anisotropic superfluidity with mixed $s$- and $p$-wave components
\cite{Hu2011b}. By imposing an external Zeeman field, novel topological
superfluid supporting zero-energy Majorana modes may also emerge \cite{Zhu2011b,Gong2011,Liu2011}.
To observe these new states of matter, it is necessarily to cool the
temperature below a threshold, which may depend critically on the
SO coupling. The purpose of this work is to determine the critical
temperature of trapped atomic Bose gases with Rashba type SO coupling.
We focus on an ideal, non-interacting Bose gas, since the critical
temperature is less affected by weak interatomic interactions \cite{Giorgini1996}.

Theoretically, the critical temperature of a homogeneous Bose gas
is greatly suppressed by the Rashba SO coupling, as the low-energy
density of states (DOS) is dramatically modified \cite{Zhai2011,Vyasanakere2011}.
In three dimensions (3D) without Rashba SO coupling, the low-energy
DOS $\rho(E)$ vanishes as $\sqrt{E}$. As a result, the number of
total particles occupied at finite energy levels, given by $N(T)=\int_{0}^{\infty}dE\rho(E)/(e^{E/k_{B}T}-1)$,
saturates at finite temperature $T$ \cite{Dalfovo1999}. This leads
to the well-known macroscopic occupation of the ground state, i.e.,
the formation of a BEC. In the presence of Rashba SO coupling, however,
the low-energy DOS becomes a constant (see Appendix A) \cite{Zhai2011,Vyasanakere2011},
reminiscent of a two-dimensional (2D) system. The thermal occupation
$N(T)$ can be logarithmically divergent. The critical temperature
is therefore precisely zero, ruling out the possibility of BEC at
any finite temperatures \cite{Dalfovo1999}.

In this paper, we show that in the presence of a harmonic trap the
Rashba SO coupling does not destroy the BEC at finite temperatures,
as the thermal occupation $N(T)$ remains finite. Actually, the critical
temperature is not affected by the Rashba SO coupling in the thermodynamic
limit where the number of particles $N$ becomes infinitely large.
This is because the occupation of low-energy states, modified by the
SO coupling, becomes negligible as $N\rightarrow\infty$. However,
in the experimentally relevant situation in which numbers of particles
range from a few thousand to a few million, we find a significant
decrease of the critical temperature and of the condensate fraction.
In particular, at a sufficiently large Rashba SO coupling strength,
the critical temperature scales like $N^{2/5}$ and $N^{2/3}$ in
three and two dimensions, respectively, in sharp contrast to the scaling
of $N^{1/3}$ and $N^{1/2}$ without SO coupling \cite{Dalfovo1999,Ketterle1996,Haugerud1997,Balaz2010}.
We derive these results either by summing discrete energy levels for
small number of particles or by using a continuous DOS under the semi-classical
assumption that the level spacing is negligible compared to the temperature.
The former approach also enables the investigation of finite-size
correction to the critical temperature.

The paper is structured as follows. In the next section (Sec. II),
we introduce the theoretical model for Rashba spin-orbit coupled ideal
Bose gases in harmonic traps and solve the single-particle energy
spectrum. In Sec. III, we present the 2D and 3D DOS with or without
the continuous spectrum approximation. The critical temperature and
condensate fraction are then calculated in Sec. IV for both 2D and
3D cases. Next, the finite size effect in 3D is discussed in Sec.
V. Finally, Sec. VI is devoted to conclusions. The calculation of
the DOS of a homogeneous 3D Rashba SO coupled system is given in the
Appendix A.

\section{Model Hamiltonian and single-particle energy spectrum}

We consider a two-component (spin-1/2) Bose gas in 2D or 3D harmonic
traps, $V_{2D}(r_{\perp})=M\omega_{\perp}^{2}(x^{2}+y^{2})/2\equiv M\omega_{\perp}^{2}r_{\perp}^{2}/2$
or $V_{3D}(r_{\perp},z)=M(\omega_{\perp}^{2}r_{\perp}^{2}+\omega_{z}^{2}z^{2})/2$,
with a Rashba SO coupling ${\cal V}_{SO}=-i\lambda_{R}(\hat{\sigma}_{x}\partial_{y}-\hat{\sigma}_{y}\partial_{x})$
in the $x-y$ plane, where $\lambda_{R}$ is the Rashba SO coupling
strength and $\hat{\sigma}_{x}$, $\hat{\sigma}_{y}$, and $\hat{\sigma}_{z}$
are the $2\times2$ Pauli matrices for pseudo-spin. The model Hamiltonian
for single-particle is described by, 
\begin{equation}
{\cal H}_{S}=\left[\begin{array}{cc}
-\hbar^{2}\nabla^{2}/2M+V_{T} & -i\lambda_{R}(\partial_{y}+i\partial_{x})\\
-i\lambda_{R}(\partial_{y}-i\partial_{x}) & -\hbar^{2}\nabla^{2}/2M+V_{T}
\end{array}\right]
\end{equation}
 where the trapping potential $V_{T}({\bf r}_{\perp})=V_{2D}(r_{\perp})$
in 2D and $V_{T}({\bf r})=V_{3D}(r_{\perp},z)$ in 3D harmonic traps.
The characteristic length scales of harmonic traps in the $x-y$ plane
and $z$-direction are given by, $a_{\perp}=\sqrt{\hbar/(M\omega_{\perp})}$
and $a_{z}=\sqrt{\hbar/(M\omega_{z})}$, respectively. For the SO
coupling, we take a dimensionless coupling strength $\lambda_{SO}\equiv\lambda_{R}Ma_{\perp}/\hbar^{2}$.

In the 2D case, it is convenient to use polar coordinates ${\bf r}_{\perp}=(r_{\perp},\varphi)$,
in which $-i(\partial_{y}\pm i\partial_{x})=e^{\mp i\varphi}[\pm\partial/\partial r_{\perp}-(i/r_{\perp})\partial/\partial\varphi]$.
As the harmonic potential is isotropic, the single-particle wave-function
has a well-defined azimuthal angular momentum $l_{z}=m$ and takes
the form, 
\begin{equation}
\phi_{m}({\bf r}_{\perp})=\left[\begin{array}{c}
\phi_{\uparrow}(r_{\perp})\\
\phi_{\downarrow}(r_{\perp})e^{i\varphi}
\end{array}\right]\frac{e^{im\varphi}}{\sqrt{2\pi}},
\end{equation}
 which preserves the total angular momentum $j_{z}=l_{z}+s_{z}=m+1/2$.
The Schrödinger equation for $\phi_{\uparrow}(r_{\perp})$ and $\phi_{\downarrow}(r_{\perp})$
therefore becomes,

\begin{widetext}

\begin{equation}
\left[\begin{array}{cc}
{\cal H}_{m} & \lambda_{R}\left[\partial/\partial r_{\perp}+\left(m+1\right)/r_{\perp}\right]\\
\lambda_{R}\left(-\partial/\partial r_{\perp}+m/r_{\perp}\right) & {\cal H}_{m+1}
\end{array}\right]\left[\begin{array}{c}
\phi_{\uparrow}\\
\phi_{\downarrow}
\end{array}\right]=E_{nm}\left[\begin{array}{c}
\phi_{\uparrow}\\
\phi_{\downarrow}
\end{array}\right],
\end{equation}

\end{widetext}where ${\cal H}_{m}\equiv-[\hbar^{2}/(2M)][d/dr_{\perp}{}^{2}+(1/r_{\perp})d/dr_{\perp}-m^{2}/r_{\perp}{}^{2}]+M\omega_{\perp}^{2}r_{\perp}{}^{2}/2$
is the 2D harmonic oscillator. We have denoted the energy level as
$E_{nm}$, with $n=(0,1,2...)$ being the good quantum number in the
transverse (radial) direction. Each energy level is two-fold degenerate,
as a result of the time-reversal symmetry satisfied by the single-particle
model Hamiltonian (Kramer's degeneracy). Any state $\phi({\bf r}_{\perp})=[\phi_{\uparrow}({\bf r}_{\perp}),\phi_{\downarrow}({\bf r}_{\perp})]^{T}$
is degenerate with its time-reversal partner ${\cal T}\phi({\bf r}_{\perp})\equiv(i\sigma_{y}{\cal C})\phi({\bf r}_{\perp})=$
$[\phi_{\downarrow}^{*}({\bf r}_{\perp}),-\phi_{\uparrow}^{*}({\bf r}_{\perp})]^{T}$,
where ${\cal C}$ is the complex conjugate operator. Therefore, we
may restrict the quantum number $m$ to be non-negative integers,
as a state with negative $m$ can always be treated as the time-reversal
partner of a state with $m\geq0$. To solve numerically the single-particle
spectrum, we expand the wave-function using the basis of 2D harmonic
oscillator, 
\begin{eqnarray}
\phi_{\uparrow}(r_{\perp}) & = & \sum_{k=0}^{\infty}A_{\uparrow k}R_{km}\left(r_{\perp}\right),\\
\phi_{\downarrow}(r_{\perp}) & = & \sum_{k=0}^{\infty}A_{\downarrow k}R_{km+1}\left(r_{\perp}\right),
\end{eqnarray}
 where 
\begin{equation}
R_{km}=\frac{1}{a_{\perp}}\left[\frac{2k!}{\left(k+\left|m\right|\right)!}\right]^{1/2}\left(\frac{r_{\perp}}{a_{\perp}}\right)^{\left|m\right|}e^{-\frac{r_{\perp}^{2}}{2a_{\perp}^{2}}}{\cal L}_{k}^{\left|m\right|}(\frac{r_{\perp}^{2}}{a_{\perp}^{2}})
\end{equation}
 is the radial wave-function of ${\cal H}_{m}$ with energy $(2k+\left|m\right|+1)\hbar\omega_{\perp}$
and ${\cal L}_{k}^{\left|m\right|}$ is the associated Legendre polynomial.
This leads to the following secular equation, 
\begin{equation}
\left[\begin{array}{cc}
{\cal H}_{m} & {\cal M}^{T}\\
{\cal M} & {\cal H}_{m+1}
\end{array}\right]\left[\begin{array}{c}
A_{\uparrow}\\
A_{\downarrow}
\end{array}\right]=E_{nm}\left[\begin{array}{c}
A_{\uparrow}\\
A_{\downarrow}
\end{array}\right],\label{spMatrix}
\end{equation}
 where the vectors $A_{\uparrow}$ and $A_{\downarrow}$ denote collectively
the expanding coefficients $\{A_{\uparrow k}\}$ and $\{A_{\downarrow k}\}$,
and the elements of matrices ($m\geq0$) are given by, 
\begin{eqnarray}
\left({\cal H}_{m}\right)_{kk^{\prime}} & = & \left(2k+m+1\right)\delta_{kk^{\prime}}\hbar\omega_{\perp},\\
{\cal M}_{kk^{\prime}} & = & \lambda_{SO}\left(\sqrt{k^{\prime}+m+1}\delta_{kk^{\prime}}+\sqrt{k^{\prime}}\delta_{kk^{\prime}-1}\right)\hbar\omega_{\perp}.
\end{eqnarray}
 Diagonalization of the secular matrix Eq. (\ref{spMatrix}) leads
to the single-particle spectrum and single-particle wave-functions.
In numerical calculations, it is necessary to truncate the radial
quantum number $k$ of the 2D harmonic oscillator, by restricting
$k<k_{\max}$. For $\lambda_{SO}\leq20$, we find that $k_{\max}=256$
is already sufficiently large to have an accurate energy spectrum.
With this cut-off, the dimension of the secular matrix in Eq. (\ref{spMatrix})
is $2k_{\max}=512$.

\begin{figure}[htp]

\begin{centering}
\includegraphics[clip,width=0.48\textwidth]{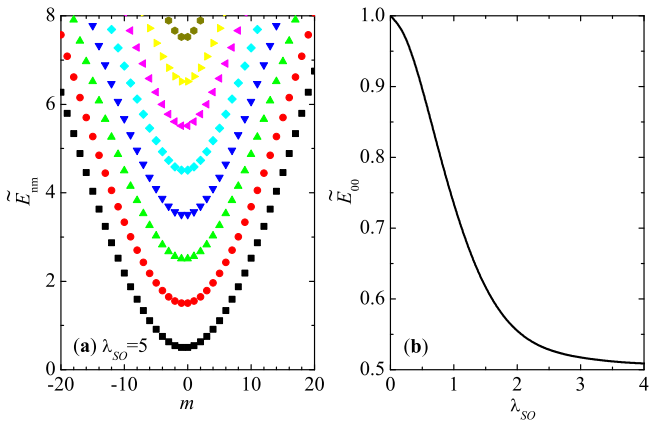} 
\par\end{centering}

\caption{(color online). (a) Single-particle energy spectrum $\tilde{E}_{nm}=E_{nm}+(\lambda_{SO}^{2}/2)\hbar\omega_{\perp}$
at $\lambda_{SO}=5$, measured in reference to the semi-classical
zero-point energy $-(\lambda_{SO}^{2}/2)\hbar\omega_{\perp}$. (b)
Ground state single-particle energy, $\tilde{E}_{00}=E_{00}+(\lambda_{SO}^{2}/2)\hbar\omega_{\perp}$,
as a function of the dimensionless Rashba SO coupling constant. The
energy is plotted in units of $\hbar\omega_{\perp}$.}

\label{fig1} 
\end{figure}

In Fig. 1a, we present the single-particle energy spectrum at $\lambda_{SO}=5$.
The ground state single-particle energy is plotted in Fig. 1b as a
function of the dimensionless SO coupling constant. In reference to
the semi-classical zero-point energy $E_{0}^{sc}\equiv-(\lambda_{SO}^{2}/2)\hbar\omega_{\perp}$,
the ground state energy decrease from $\hbar\omega_{\perp}$ to $\hbar\omega_{\perp}/2$,
when the Rashba SO coupling strength $\lambda_{SO}$ becomes sufficiently
large. In that limit (i.e., $\lambda_{SO}>5$), the low-lying energy
spectrum becomes fairly flat, with a dispersion that is well approximated
by \cite{Hu2011a,Sinha2011}, 
\begin{equation}
E_{nm}\simeq\left[-\frac{\lambda_{SO}^{2}}{2}+(n+\frac{1}{2})+\frac{m\left(m+1\right)}{2\lambda_{SO}^{2}}\right]\hbar\omega_{\perp}.\label{dispersionLargeSO}
\end{equation}

In 3D, because the motions in $xy$-plane and $z$-direction are decoupled,
the single-particle energy spectrum is given by, 
\begin{equation}
E_{nmn_{z}}=E_{nm}+\left(n_{z}+\frac{1}{2}\right)\hbar\omega_{z},
\end{equation}
 where $n_{z}=0,1,2...$ is a good quantum number for the axial motion.

At finite temperature $T$, the total number of particles is given,
in the grand-canonical ensembles, by the sum 
\begin{equation}
N=\sum_{n,m=0}^{\infty}\frac{2}{\exp\left[\left(E_{nm}-\mu\right)/k_{B}T\right]-1}\label{discreteNumSum2D}
\end{equation}
 in 2D and by the sum 
\begin{equation}
N=\sum_{n,m,n_{z}=0}^{\infty}\frac{2}{\exp\left[\left(E_{nmn_{z}}-\mu\right)/k_{B}T\right]-1}\label{discreteNumSum3D}
\end{equation}
 in 3D, where $\mu$ is the chemical potential and the factor of $2$
arises from the Kramer's degeneracy. The sum can be rewritten as an
integral over the energy, in the unified form, 
\begin{equation}
N=\int_{-\infty}^{+\infty}dE\frac{\rho\left(E\right)}{\exp\left[\left(E-\mu\right)/k_{B}T\right]-1},
\end{equation}
 with the DOS $\rho\left(E\right)$ given by 
\begin{equation}
\rho_{2D}\left(E\right)=2\sum_{n,m=0}^{\infty}\delta\left(E-E_{nm}\right)\label{discreteDOS2D}
\end{equation}
 and 
\begin{equation}
\rho_{2D}\left(E\right)=2\sum_{n,m,n_{z}=0}^{\infty}\delta\left(E-E_{nmn_{z}}\right),\label{discreteDOS3D}
\end{equation}
 in 2D and 3D, respectively.

For given small numbers of particles $N$, we can calculate the low-lying
energy levels and then sum explicitly the number equations, Eqs. (\ref{discreteNumSum2D})
and (\ref{discreteNumSum3D}). Once the chemical potential is determined
at a given temperature $T$, we calculate the occupation of the ground
state, 
\begin{equation}
N_{0}=\frac{2}{\exp\left[\left(E_{0}-\mu\right)/k_{B}T\right]-1},\label{discreteN0}
\end{equation}
 where the single-particle ground state energy $E_{0}=E_{00}$ in
2D and $E_{0}=E_{00}+\hbar\omega_{z}/2$ in 3D. The BEC transition
temperature $T_{c}$ can be determined from $d^{2}N_{0}/dT^{2}$,
which exhibits a maximum at $T_{c}$ \cite{Bergeman2000}.

\section{Semi-classical density of states}

For large numbers of particles, it is useful to consider a semi-classical
approximation by using continuous energy spectrum \cite{Dalfovo1999}.
The level spacing, typical of $\hbar\omega_{\perp}$ or $\hbar\omega_{z}$,
is assumed to be negligibly small, compared with the thermal energy
$k_{B}T$. Thus, the relevant excitation energies, contributing to
the sum in Eqs. (\ref{discreteNumSum2D}) and (\ref{discreteNumSum3D}),
are much larger than the level spacing. The accuracy of the semi-classical
approximation can be tested a posteriori by comparing the semi-classical
result with the numerical discrete summation.

\subsection{2D density of states}

In 2D, the semi-classical DOS can be written as, 
\begin{equation}
\rho_{2D}^{sc}\left(E\right)=\sum_{s=\pm}\int\frac{d{\bf r}_{\perp}d{\bf k}_{\perp}}{\left(2\pi\right)^{2}}\delta\left[E_{{\bf k}s}({\bf r}_{\perp})-E\right],
\end{equation}
 where $E_{{\bf k}s}({\bf r}_{\perp})=\hbar^{2}k_{\perp}^{2}/(2M)+s\lambda_{R}k_{\perp}+M\omega_{\perp}^{2}r_{\perp}^{2}/2$
is the semi-classical energy in phase space (${\bf r}_{\perp},{\bf k}_{\perp}$).
Because of the Rashba SO coupling, the semi-classical energy splits
into two helicity branches as indicated by $s=\pm$ (see Appendix
A). By integrating out the spatial degree of freedom, we obtain that,
\begin{equation}
\hbar\omega_{\perp}\rho_{2D}^{sc}\left(E\right)=\sum_{s=\pm}\int\limits _{0}^{\infty}\tilde{k}_{\perp}d\tilde{k}_{\perp}\Theta\left[\frac{\tilde{E}}{\hbar\omega_{\perp}}-\frac{\left(\tilde{k}_{\perp}+s\lambda_{SO}\right)^{2}}{2}\right],\label{semiclassicalDOS2D}
\end{equation}
where $\tilde{k}_{\perp}\equiv k_{\perp}a_{\perp}$ is the dimensionless
wave-vector, $\tilde{E}\equiv E+(\lambda_{SO}^{2}/2)\hbar\omega_{\perp}$
is the energy measured in reference to the semi-classical zero-point
energy $E_{0}^{sc}\equiv-(\lambda_{SO}^{2}/2)\hbar\omega_{\perp}$,
and $\Theta(\cdot)$ is the Heaviside step function. The integration
over the wave-vector can be calculated explicitly as well. We finally
arrive at, \begin{widetext} 
\begin{equation}
\hbar\omega_{\perp}\rho_{2D}^{sc}\left(E\right)=\left\{ \begin{array}{ll}
0, & \left(E<E_{0}^{sc}\right);\\
2\lambda_{SO}\left[2E/\left(\hbar\omega_{\perp}\right)+\lambda_{SO}^{2}\right]^{1/2}, & (E_{0}^{sc}\leq E<0);\\
2E/\left(\hbar\omega_{\perp}\right)+2\lambda_{SO}^{2}, & (E\geq0).
\end{array}\right.
\end{equation}
 In the absence of Rashba SO coupling ($\lambda_{SO}=0$), we recover
the usual expression for the 2D DOS in harmonic traps, $\rho_{2D}^{sc}\left(E\right)=2E/(\hbar\omega_{\perp})^{2}\Theta(E)$,
for a two-component system \cite{Dalfovo1999}.

\subsection{3D density of states}

Likewise, we calculate the semi-classical DOS in 3D, which is given
by, 
\begin{equation}
\rho_{3D}^{sc}\left(E\right)=\sum_{s=\pm}\int\frac{d{\bf r}d{\bf k}}{\left(2\pi\right)^{3}}\delta\left[E_{{\bf k}s}({\bf r})-E\right],
\end{equation}
 where the semi-classical energy now takes the form $E_{{\bf k}s}({\bf r})=\hbar^{2}k_{\perp}^{2}/(2M)+s\lambda_{R}k_{\perp}+\hbar^{2}k_{z}^{2}/(2M)+M(\omega_{\perp}^{2}r_{\perp}^{2}+\omega_{z}^{2}z^{2})/2$.
The integration over ${\bf r}$ and $k_{z}$ can be done by introducing
a new variable $t^{2}=\hbar^{2}k_{z}^{2}/(2M)+M(\omega_{\perp}^{2}r_{\perp}^{2}+\omega_{z}^{2}z^{2})/2$
and by converting the variables of integration $d{\bf r}d{\bf k}$
to $d{\bf t}d{\bf k}_{\perp}$. This leads to, 
\begin{equation}
\hbar\omega_{z}\rho_{3D}^{sc}\left(E\right)=\sum_{s=\pm}\int\limits _{0}^{\infty}\tilde{k}_{\perp}d\tilde{k}_{\perp}\left[\frac{\tilde{E}}{\hbar\omega_{\perp}}-\frac{\left(\tilde{k}_{\perp}+s\lambda_{SO}\right)^{2}}{2}\right]\Theta\left[\frac{\tilde{E}}{\hbar\omega_{\perp}}-\frac{\left(\tilde{k}_{\perp}+s\lambda_{SO}\right)^{2}}{2}\right].\label{semiclassicalDOS3D}
\end{equation}
 By explicitly integrating out $\tilde{k}_{\perp}$, we obtain, 
\begin{equation}
\hbar\omega_{z}\rho_{3D}^{sc}\left(E\right)=\left\{ \begin{array}{ll}
0, & \left(E<E_{0}^{sc}\right);\\
\left(2\lambda_{SO}/3\right)\left[2E/\left(\hbar\omega_{\perp}\right)+\lambda_{SO}^{2}\right]^{3/2}, & (E_{0}^{sc}\leq E<0);\\
\left[E/\left(\hbar\omega_{\perp}\right)+\lambda_{SO}^{2}\right]^{2}-\lambda_{SO}^{4}/3, & (E\geq0).
\end{array}\right.
\end{equation}
\end{widetext} In the absence of Rashba SO coupling, we recover the
expression $\rho_{3D}^{sc}\left(E\right)=E^{2}/\left(\hbar^{3}\omega_{\perp}^{2}\omega_{z}\right)\Theta(E)$
for 3D harmonic traps \cite{Dalfovo1999}.

It is easy to check that the 2D and 3D DOS is related by 
\begin{equation}
\hbar\omega_{z}\frac{d\rho_{3D}^{sc}\left(E\right)}{dE}=\rho_{2D}^{sc}\left(E\right).\label{relationDOS}
\end{equation}
 This is due to the decoupled motion in the $x-y$ plane and $z$
direction, which leads to the observation that the 3D energy spectrum
may alternatively be viewed as a collection of 2D spectra with regular
spacing $\hbar\omega_{z}$.

\subsection{Test of the semi-classical DOS}

\begin{figure}[htp]

\begin{centering}
\includegraphics[clip,width=0.4\textwidth]{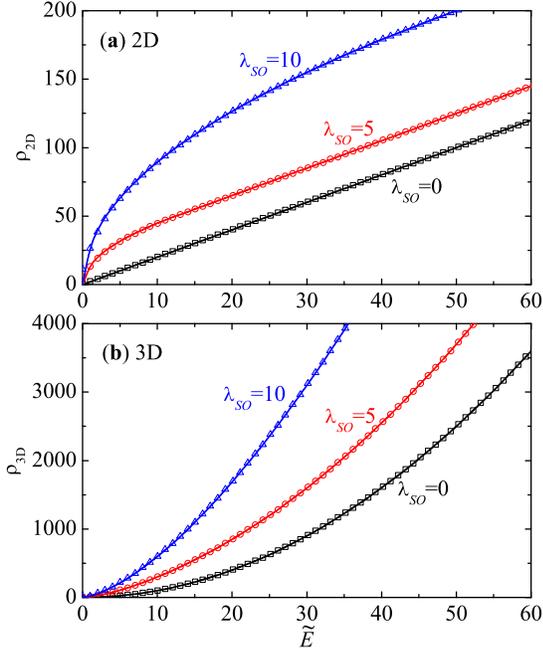} 
\par\end{centering}

\caption{(color online). The semi-classical density of states in 2D (a) and
3D (b), in units of $1/(\hbar\omega_{\perp})$ and $1/(\hbar\omega_{z})$
respectively, are shown as a function of $\tilde{E}=E+(\lambda_{SO}^{2}/2)\hbar\omega_{\perp}$
at different Rashba SO couplings (solid lines). The energy $\tilde{E}$
is in units of $\hbar\omega_{\perp}$. For comparison, the symbols
plot the results obtained by the numerical summation, see, Eqs.
(\ref{discreteDOS2D}) and (\ref{discreteDOS3D}). The simulation
of the delta-function is described in the text.}

\label{fig2} 
\end{figure}

In Fig. 2, we compare the semi-classical 2D and 3D DOS with these
obtained by summing over the discrete single-particle energy spectrum
using Eqs. (\ref{discreteDOS2D}) and (\ref{discreteDOS3D}). In the
numerical summation, we simulate the $\delta$-function $\delta(x)$
by a Lorentzen line shape with broadening $\Gamma$, $f_{\delta}(x;\Gamma)=(\Gamma/\pi)/(x^{2}+\Gamma^{2})$.
Roughly, the resulting DOS depends linearly on $\Gamma$ at $\Gamma\sim\hbar\omega_{\perp}$.
Therefore, we use 
\begin{equation}
\delta(x)=2f_{\delta}(x;\Gamma=\hbar\omega_{\perp})-f_{\delta}(x;\Gamma=2\hbar\omega_{\perp}),
\end{equation}
 as an extrapolation to the zero-broadening limit ($\Gamma=0$). We
find that the semi-classical expressions for DOS, Eqs. (\ref{semiclassicalDOS2D})
and (\ref{semiclassicalDOS3D}), works extremely well over a very
broad range for energy. The most significant discrepancy occurs at
the lowest energy level, $E\sim-(\lambda_{SO}^{2}/2)\hbar\omega_{\perp}$,
as anticipated.

\section{Critical temperature and condensate fraction}

We are now ready to calculate the critical temperature and condensate
fraction for large number of particles. With the semi-classical DOS
$\rho^{sc}\left(E\right)$, the number of particles could be rewritten
as \cite{Dalfovo1999}, 
\begin{equation}
N=N_{0}+\int_{E_{0}^{sc}}^{+\infty}dE\frac{\rho^{sc}\left(E\right)}{\exp\left[\left(E-\mu\right)/k_{B}T\right]-1},
\end{equation}
 where the ground state population $N_{0}$ is singled out and the
finite sum over the excited states in Eqs. (\ref{discreteNumSum2D})
and (\ref{discreteNumSum3D}) is replaced by an integral. Accordingly,
we have set the lower-bound of the integral to be the semi-classical
zero-point energy $E_{0}^{sc}=-(\lambda_{SO}^{2}/2)\hbar\omega_{\perp}$.
When BEC occurs, the chemical potential approaches to $E_{0}^{sc}$
from below \cite{Dalfovo1999}. The critical temperature $T_{c}$
is determined by the condition, 
\begin{equation}
N=\int_{0}^{+\infty}d\tilde{E}\frac{\rho^{sc}\left(\tilde{E}+E_{0}^{sc}\right)}{\exp\left[\tilde{E}/k_{B}T_{c}\right]-1},\label{eqTc0}
\end{equation}
 where $\tilde{E}\equiv E-E_{0}^{sc}$, and the condensate fraction
at $T<T_{c}$ can be calculated by, 
\begin{equation}
\frac{N_{0}}{N}=1-\frac{1}{N}\int_{0}^{+\infty}d\tilde{E}\frac{\rho^{sc}\left(\tilde{E}+E_{0}^{sc}\right)}{\exp\left[\tilde{E}/k_{B}T\right]-1}.
\end{equation}
 As we shall see, these equations can be conveniently solved by introducing
$\epsilon=\tilde{E}/(k_{B}T)$ and 
\begin{equation}
\alpha(T)=\lambda_{SO}\sqrt{\frac{\hbar\omega_{\perp}}{k_{B}T}}.\label{alpha}
\end{equation}

\subsection{2D}

In 2D, the equations for the critical temperature and condensate fraction
becomes,

\begin{equation}
N=\left(\frac{k_{B}T_{c}}{\hbar\omega_{\perp}}\right)^{2}{\cal I}_{2D}\left[\alpha\left(T_{c}\right)\right]
\end{equation}
 and 
\begin{equation}
\frac{N_{0}}{N}=1-\left(\frac{T}{T_{c}}\right)^{2}\frac{{\cal I}_{2D}\left[\alpha\left(T\right)\right]}{{\cal I}_{2D}\left[\alpha\left(T_{c}\right)\right]},
\end{equation}
 respectively. Here the integral ${\cal I}_{2D}\left[\alpha\right]$
takes the form, 
\begin{equation}
{\cal I}_{2D}\left[\alpha\right]=\int_{0}^{+\infty}d\epsilon\frac{\tilde{\rho}_{2D}^{sc}\left(\epsilon;\alpha\right)}{e^{\epsilon}-1},
\end{equation}
 where the dimensionless DOS $\tilde{\rho}_{2D}^{sc}\left(\epsilon;\alpha\right)$
is given by, 
\begin{equation}
\tilde{\rho}_{2D}^{sc}\left(\epsilon;\alpha\right)=\left\{ \begin{array}{ll}
0, & \left(\epsilon<0\right);\\
2\alpha\sqrt{2\epsilon}, & (0\leq\epsilon<\alpha^{2}/2);\\
2\epsilon+\alpha^{2}, & (\epsilon\geq\alpha^{2}/2).
\end{array}\right.
\end{equation}
 Therefore, 
\begin{equation}
{\cal I}_{2D}\left[\alpha\right]=\sqrt{2\pi}\alpha\zeta\left(\frac{3}{2}\right)+\int\limits _{\alpha^{2}/2}^{+\infty}d\epsilon\frac{\left(\sqrt{2\epsilon}-\alpha\right)^{2}}{e^{\epsilon}-1}.
\end{equation}
 Here $\zeta(\cdot)$ is the Riemann $\zeta$ function. ${\cal I}_{2D}\left[\alpha\right]$
depends implicitly on the temperature through the dimensionless parameter
$\alpha(T)$. It is clear from Eq. (\ref{alpha}) that for a given
SO coupling $\lambda_{SO}$, the dimensionless parameter $\alpha$
at the critical temperature $T_{c}$ always scales to zero in the
thermodynamic limit $N\rightarrow\infty$. This is understandable
as a finite SO interaction modifies only the low-lying energy states,
whose occupation becomes negligible as $N\rightarrow\infty$.

In the absence of SO coupling, ${\cal I}_{2D}[\alpha=0]=2\zeta(2)=\pi^{2}/3$,
we recover the standard results in 2D \cite{Dalfovo1999}, 
\begin{equation}
k_{B}T_{c}^{(0)}\left(\lambda_{SO}=0\right)=\frac{1}{\pi}\left(3N\right)^{1/2}\hbar\omega_{\perp}
\end{equation}
 and $N_{0}/N=1-(T/T_{c}^{(0)})^{2}$. Here, we use the superscript
``$0$'' to indicate the semi-classical result. For a large SO coupling,
${\cal I}_{2D}\left[\alpha\gg1\right]=\sqrt{2\pi}\alpha\zeta\left(3/2\right)$,
we find 
\begin{equation}
k_{B}T_{c}^{(0)}\left(\lambda_{SO}\gg1\right)=\frac{1}{\left(2\pi\right)^{1/3}}\left[\frac{N}{\lambda_{SO}\zeta\left(3/2\right)}\right]^{2/3}\hbar\omega_{\perp}\label{strongcouplingTc2D}
\end{equation}
 and $N_{0}/N=1-(T/T_{c}^{(0)})^{3/2}$. Thus, for a given number
of particles, with increasing SO coupling the dependence of 2D critical
temperature on the number of particles changes from $N^{1/2}$ to
$N^{2/3}$. Using $\alpha\gg1$, the strong-coupling limit is reached
when 
\begin{equation}
\lambda_{SO}\gg\left(2\pi\right)^{-1/8}\left[\frac{N}{\zeta\left(3/2\right)}\right]^{1/4}.
\end{equation}

\begin{figure}[htp]

\begin{centering}
\includegraphics[clip,width=0.4\textwidth]{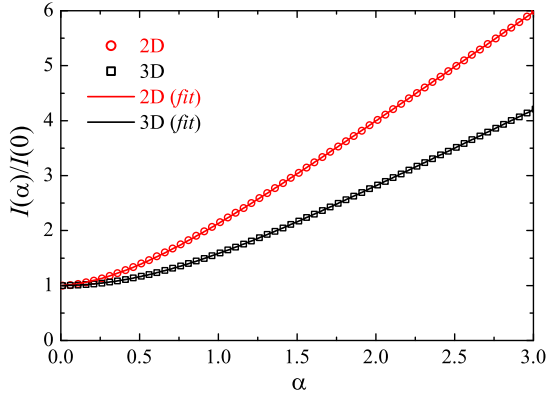} 
\par\end{centering}

\caption{(color online). The integrals ${\cal I}_{2D}$ and ${\cal I}_{3D}$
as a function of the dimensionless parameter $\alpha=\lambda_{SO}[\hbar\omega_{\perp}/(k_{B}T)]^{1/2}$
(symbols). The solid lines show the empirical fit which agrees numerically
within $0.5\%$ relative error (see the text for the empirical formalism).}

\label{fig3} 
\end{figure}

\begin{figure}[htp]

\begin{centering}
\includegraphics[clip,width=0.4\textwidth]{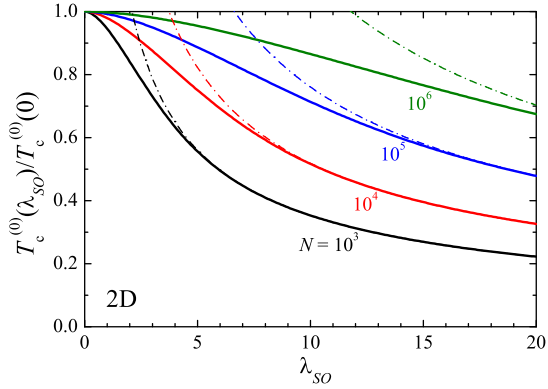} 
\par\end{centering}

\caption{(color online). 2D critical temperature as a function of the SO coupling
at different numbers of particles, as indicated. The dot-dashed lines
show the limiting behavior at large SO coupling, Eq. (\ref{strongcouplingTc2D}).}

\label{fig4} 
\end{figure}

\begin{figure}[htp]

\begin{centering}
\includegraphics[clip,width=0.4\textwidth]{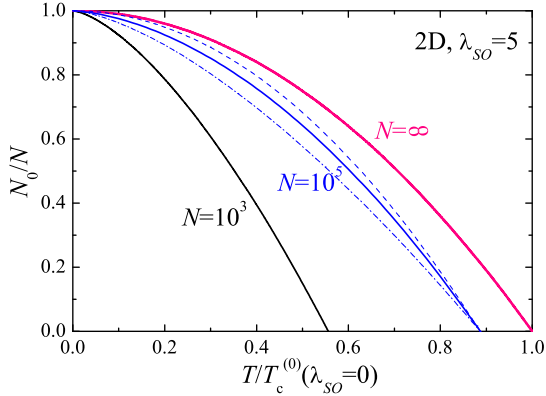} 
\par\end{centering}

\caption{(color online). 2D condensate fraction at $\lambda_{SO}=5$ and at
different numbers of particles. For the case of $N=10^{5}$, the dashed
and dot-dashed lines show respectively the strong-coupling and zero-coupling
result, $N_{0}/N=1-(T/T_{c}^{(0)})^{2}$ and $N_{0}/N=1-(T/T_{c}^{(0)})^{3/2}$.}

\label{fig5} 
\end{figure}

In Fig. 3, we show ${\cal I}_{2D}\left[\alpha\right]$ as a function
of the dimensionless parameter $\alpha$. Empirically, we find that
${\cal I}_{2D}\left[\alpha\right]\simeq\sqrt{2\pi}\alpha\zeta\left(3/2\right)+2\zeta(2)e^{-1.84\alpha-0.13\alpha^{2}}$,
within $0.5\%$ relative error. Fig. 4 reports the critical temperature
as a function of SO coupling at several numbers of particles (solid
lines). It decreases significantly at moderate SO coupling ($\lambda_{SO}\sim10$)
and number of particles (i.e., $N\sim10^{4}$). The strong-coupling
results Eq. (\ref{strongcouplingTc2D}) have also been plotted using
dot-dashed lines. Finally, in Fig. 5, we present the condensate fraction
at $\lambda_{SO}=5$ and $N=10^{3}$, $10^{5}$, and $\infty$.

\subsection{3D}

In 3D, similarly we obtain that

\begin{equation}
N\lambda=\left(\frac{k_{B}T_{c}}{\hbar\omega_{\perp}}\right)^{3}{\cal I}_{3D}\left[\alpha\left(T_{c}\right)\right]
\end{equation}
 and 
\begin{equation}
\frac{N_{0}}{N}=1-\left(\frac{T}{T_{c}}\right)^{3}\frac{{\cal I}_{3D}\left[\alpha\left(T\right)\right]}{{\cal I}_{3D}\left[\alpha\left(T_{c}\right)\right]},
\end{equation}
 where $\lambda=\omega_{z}/\omega_{\perp}$ is the aspect ratio of
the harmonic trap, the integral ${\cal I}_{3D}\left[\alpha\right]$
is given by, 
\begin{equation}
{\cal I}_{3D}\left[\alpha\right]=\int_{0}^{+\infty}d\epsilon\frac{\tilde{\rho}_{3D}^{sc}\left(\epsilon;\alpha\right)}{e^{\epsilon}-1},
\end{equation}
 and the dimensionless DOS $\tilde{\rho}_{3D}^{sc}\left(\epsilon;\alpha\right)$
is, 
\begin{equation}
\tilde{\rho}_{3D}^{sc}\left(\epsilon;\alpha\right)=\left\{ \begin{array}{ll}
0, & \left(\epsilon<0\right);\\
(4\sqrt{2}\alpha/3)\epsilon^{3/2}, & (0\leq\epsilon<\alpha^{2}/2);\\
\epsilon^{2}+\epsilon\alpha^{2}-\alpha^{4}/12, & (\epsilon\geq\alpha^{2}/2).
\end{array}\right.
\end{equation}
 Explicitly, we find that
\begin{equation}
{\cal I}_{3D}\left[\alpha\right]=\sqrt{2\pi}\alpha\zeta\left(\frac{5}{2}\right)+\int\limits _{\alpha^{2}/2}^{+\infty}d\epsilon\frac{h\left(\epsilon\right)}{e^{\epsilon}-1},
\end{equation}
where $h\left(\epsilon\right)=\epsilon^{2}+\epsilon\alpha^{2}-\alpha^{4}/12-(4\sqrt{2}\alpha/3)\epsilon^{3/2}$.
We plot ${\cal I}_{3D}\left[\alpha\right]$$ $ in Fig. 3, together
with an empirical fit, ${\cal I}_{3D}\left[\alpha\right]=\sqrt{2\pi}\alpha\zeta\left(5/2\right)+2\zeta(3)e^{-1.40\alpha-0.30\alpha^{2}}$.

At $\lambda_{SO}=0$ where ${\cal I}_{3D}[\alpha=0]=2\zeta(3)$, we
obtain 
\begin{equation}
k_{B}T_{c}^{(0)}\left(\lambda_{SO}=0\right)=\left[\frac{N\lambda}{2\zeta(3)}\right]^{1/3}\hbar\omega_{\perp}
\end{equation}
 and $N_{0}/N=1-(T/T_{c}^{(0)})^{3}$, recovering the well-known 3D
result for a trapped spin-1/2 Bose gas \cite{Dalfovo1999}. In the
limit of large SO coupling where ${\cal I}_{3D}\left[\alpha\gg1\right]=\sqrt{2\pi}\alpha\zeta\left(5/2\right)$,
we find instead 
\begin{equation}
k_{B}T_{c}^{(0)}\left(\lambda_{SO}\gg1\right)=\frac{1}{\left(2\pi\right)^{1/5}}\left[\frac{N\lambda}{\lambda_{SO}\zeta\left(5/2\right)}\right]^{2/5}\hbar\omega_{\perp}\label{strongcouplingTc3D}
\end{equation}
 and $N_{0}/N=1-(T/T_{c}^{(0)})^{5/2}$. Thus, for given $N$, with
increasing SO coupling the power-law dependence of 3D critical temperature
on number of particles changes from $N^{1/3}$ to $N^{2/5}$. We estimate
that the strong-coupling result is applicable if 
\begin{equation}
\lambda_{SO}\gg\left(2\pi\right)^{-1/12}\left[\frac{N\lambda}{\zeta\left(5/2\right)}\right]^{1/6}.
\end{equation}

\begin{figure}[htp]

\begin{centering}
\includegraphics[clip,width=0.48\textwidth]{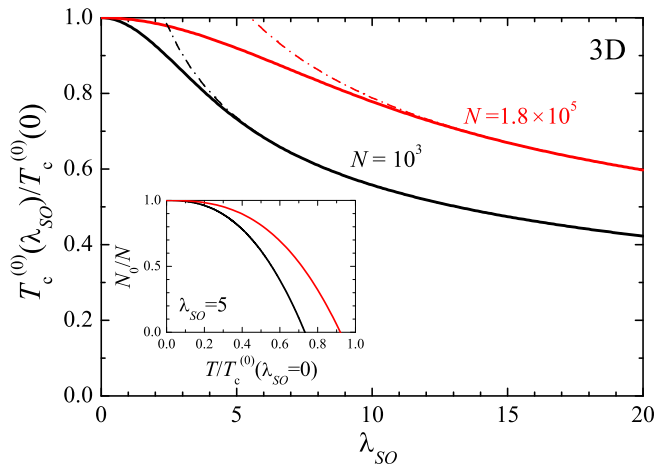} 
\par\end{centering}

\caption{(color online). 3D critical temperature as a function of the SO coupling
strength at $N=10^{3}$ and $N=1.8\times10^{5}$. The aspect ratio
of the harmonic trap is $\lambda=\omega_{z}/\omega_{\perp}=\sqrt{8}$.
The dot-dashed lines show the critical temperature in the strong-coupling
limit, Eq. (\ref{strongcouplingTc3D}). The inset reports the condensate
fraction at $\lambda_{SO}=5$.}

\label{fig6} 
\end{figure}

In Fig. 6, we report the effect of the SO coupling on 3D critical
temperature. To make a connection with the NIST experiment \cite{Lin2011},
we have used a realistic aspect ratio of the trapping potential and
number of particles, $\lambda=\sqrt{8}$ and $N=1.8\times10^{5}$.
We also consider the case with a small number of particles $N=10^{3}$.
At the typical SO coupling strength $\lambda_{SO}\sim10$ \cite{Lin2011},
the reduction of the critical temperature is about $20\%$, which
is in reach of current experiments. The inset shows the condensate
fraction at $\lambda_{SO}=5$.

\section{Finite size correction to $T_{c}$ in 3D}

We now turn to consider the finite size correction to the semi-classical
results, which arises from the discreteness of the single-particle
energy spectrum \cite{Ketterle1996,Haugerud1997}. The semi-classical
results are obtained using the semi-classical approximation for the
excited states and setting the chemical potential to the semi-classical
zero-point energy $E_{0}^{sc}$. To the leading order, the finite
size correction can be included by still employing the semi-classical
description for the excited states, while keeping the quantum value
$\mu=E_{0}$ for the chemical potential at the transition \cite{Giorgini1996}.
Here, $E_{0}>E_{0}^{sc}$ is the single-particle energy of the ground
state. It is $E_{00}$ in 2D and $E_{00}+\hbar\omega_{z}/2$ in 3D;
see, for example, Fig. (1b) for $E_{00}$ as a function of the SO
coupling strength. The discreteness of the excited energy spectrum
gives rise to higher-order finite size corrections. In the following,
we focus on the finite size correction to the 3D critical temperature.

Using the quantum value $\mu=E_{0}$ for the chemical potential, the
3D critical temperature is determined by, 
\begin{eqnarray}
N & = & \int_{E_{0}}^{+\infty}dE\frac{\rho_{3D}^{sc}\left(E\right)}{\exp\left[\left(E-E_{0}\right)/k_{B}T_{c}\right]-1},\\
 & = & \int_{0}^{+\infty}dE\frac{\rho_{3D}^{sc}\left(\tilde{E}+E_{0}^{sc}+\Delta E\right)}{\exp\left[\tilde{E}/k_{B}T_{c}\right]-1},
\end{eqnarray}
 where in the second line we have introduced $\tilde{E}=E-E_{0}$
and $\Delta E=E_{0}-E_{0}^{sc}>0$. Compared with Eq. (\ref{eqTc0}),
the 3D DOS is slightly up-shifted by an amount $\Delta E$. As $\Delta E\sim\hbar\omega_{\perp}$
is the smallest energy scale, using Eq. (\ref{relationDOS}) we may
write $\rho_{3D}^{sc}(\tilde{E}+E_{0}^{sc}+\Delta E)\simeq\rho_{3D}^{sc}(\tilde{E}+E_{0}^{sc})+(\Delta E/\hbar\omega_{z})\rho_{2D}^{sc}(\tilde{E}+E_{0}^{sc})$.
Therefore, using the integrals ${\cal I}_{2D}$ and ${\cal I}_{3D}$
the equation for the critical temperature is given by 
\begin{equation}
N\lambda=\left(\frac{k_{B}T_{c}}{\hbar\omega_{\perp}}\right)^{3}\left[{\cal I}_{3D}\left[\alpha\left(T_{c}\right)\right]+\frac{\Delta E}{k_{B}T_{c}}{\cal I}_{2D}\left[\alpha\left(T_{c}\right)\right]\right].\label{eqTc}
\end{equation}
 In the absence of the SO coupling, ${\cal I}_{2D}=2\zeta(2)$, ${\cal I}_{3D}=2\zeta(3)$
and $\Delta E=\hbar\omega_{\perp}+\hbar\omega_{z}/2$, it is easy
to verify that the transition temperature $T_{c}$ is given by the
law, 
\begin{equation}
\frac{T_{c}}{T_{c}^{0}}\simeq1-\frac{\zeta(2)}{\left[2\zeta(3)\right]^{2/3}}N^{-1/3}\frac{\left(2\hbar\omega_{\perp}+\hbar\omega_{z}\right)/3}{\left(\omega_{\perp}^{2}\omega_{z}\right)^{1/3}},
\end{equation}
 which is known in the literature \cite{Giorgini1996,Ketterle1996,Haugerud1997}.

\begin{figure}[htp]

\begin{centering}
\includegraphics[clip,width=0.48\textwidth]{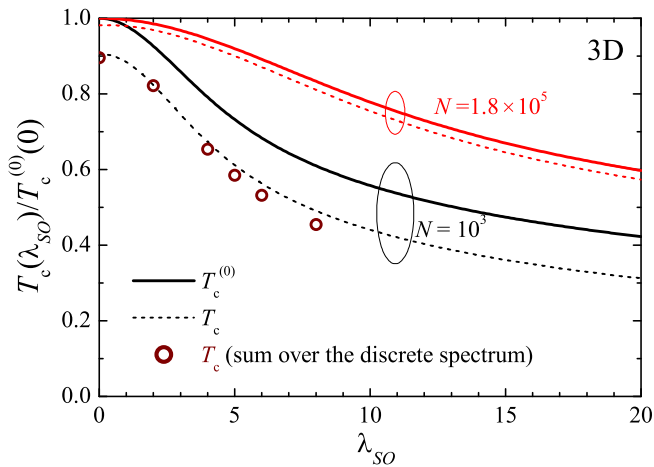} 
\par\end{centering}

\caption{(color online). 3D transition temperature as a function of the SO
coupling. The solid lines show the semi-classical predictions and
the dashed line gives the results with inclusion of the leading finite
size correction. The empty circles are calculated using the numerical
summation for $N_{0}$ with discrete energy spectrum, i.e. Eq. (\ref{discreteN0}).
The critical temperature is then determined from the peak position
of $d^{2}N_{0}/dT^{2}$ \cite{Bergeman2000}.}

\label{fig7} 
\end{figure}

In Fig. 7, we report the 3D transition temperature with the leading
finite size correction, as shown by dashed lines. We find a sizable
correction at small number of particles (i.e., $N=10^{3}$). For the
experimentally realistic number of particles, i.e., $N=1.8\times10^{5}$,
however, the correction becomes mild. As a benchmark to our analytic
treatment for $T_{c}$, we also show by symbols the critical temperature
for small number of particles, calculated by the discrete sum for
the ground state population $N_{0}$, Eq. (\ref{discreteN0}). At
relatively small SO coupling (i.e., $\lambda_{SO}<5$), our analytic
treatment works very well. However, for large SO coupling, the single-particle
level splitting between the ground state and the first excited state
becomes increasing small. We then may have to take into account the
discreteness of the low-lying excited energy levels.

\section{Conclusions}

In summary, we have investigated the critical temperature and condensate
fraction of a harmonically trapped ideal Bose gas in the presence
of Rashba spin-orbit coupling, by using either the exact numerical
summation for small number of particles or the analytic semi-classical
approach for large number of particles. The leading finite size correction
to the semi-classical approximation has also been considered. We have
found pronounced effect of the Rashba SO coupling. For the experimentally
realistic number of particles ($N\sim10^{5}$) \cite{Lin2011}, the
critical temperature is reduced by more than 20\% in magnitude at
a moderate SO coupling. This reduction is readily observable in current
experiments. Moreover, in the limit of strong SO coupling, the critical
temperature scales as $N^{2/5}$ and $N^{2/3}$ in three and two dimensions,
respectively, which should be contrasted with the scaling law of $N^{1/3}$
and $N^{1/2}$ in the absence SO coupling. Our investigation of critical
temperature can be easily extended to include a weak repulsive interaction,
by using mean-field Hartree-Fock theory \cite{Giorgini1996}.

\section*{Acknowledgments}

This work was supported by the ARC Discovery Project (Grant No. DP0984522
and DP0984637) and NFRP-China (Grant No. 2011CB921502).

\appendix
%dummy comment inserted by tex2lyx to ensure that this paragraph is not empty

\section{Density of states of a 3D homogeneous SO coupled system}

In free space, the single-particle Hamiltonian with Rashba SO coupling,
\begin{equation}
{\cal H}_{S}=\left[\begin{array}{cc}
-\hbar^{2}\nabla^{2}/2M & -i\lambda_{R}(\partial_{y}+i\partial_{x})\\
-i\lambda_{R}(\partial_{y}-i\partial_{x}) & -\hbar^{2}\nabla^{2}/2M
\end{array}\right],
\end{equation}
 has the dispersion, 
\begin{equation}
E_{{\bf k}s}=\frac{\hbar^{2}k_{z}^{2}}{2M}+\frac{\hbar^{2}k_{\perp}^{2}}{2M}+s\lambda_{R}k_{\perp}.
\end{equation}
 Here $s=\pm$ denotes the two helicity branches. The DOS, given by
$\rho(E)=(1/V)\sum_{{\bf k}}[\delta(E_{{\bf k}+}-E)+\delta(E_{{\bf k}-}-E)]$,
can be calculated analytically. We find that, \begin{widetext}
\begin{equation}
\rho\left(E\right)=\frac{M^{2}\lambda_{R}}{\hbar^{4}}\left\{ \begin{array}{ll}
0, & \left(E<-E_{R}/2\right);\\
\pi/2, & (-E_{R}/2\leq E<0);\\
\sqrt{2E/E_{R}}+\left[\pi/2-\arctan\sqrt{2E/E_{R}}\right], & (E\geq0).
\end{array}\right.
\end{equation}
 where $E_{R}\equiv M\lambda_{R}^{2}/\hbar^{2}$ is the characteristic
energy related to the SO coupling. This result was reported by Hui
Zhai in Ref. \cite{Zhai2011} (see for example, their Fig. 2b). By
introducing a Fermi wave-vector $k_{F}=(3\pi^{2}N/V)^{1/3}$, Fermi
energy $E_{F}=\hbar^{2}k_{F}^{2}/(2M)$, and dimensionless SO coupling
strength $\lambda_{eff}=M^{2}\lambda_{R}/(\hbar^{2}k_{F})$, the DOS
can be written as, 
\begin{equation}
\rho\left(E\right)=\frac{Mk_{F}}{\hbar^{2}}\left\{ \begin{array}{ll}
0, & \left(E<-\lambda_{eff}^{2}\right);\\
\lambda_{eff}\pi/2, & (-\lambda_{eff}^{2}\leq E<0);\\
\sqrt{E/E_{F}}+\lambda_{eff}\left[\pi/2-\arctan\sqrt{E/\left(\lambda_{eff}^{2}E_{F}\right)}\right], & (E\geq0).
\end{array}\right..
\end{equation}
\end{widetext} We show in Fig. 8 the DOS at different SO coupling
strengths.

\begin{figure}[htp]
\begin{centering}
\includegraphics[clip,width=0.4\textwidth]{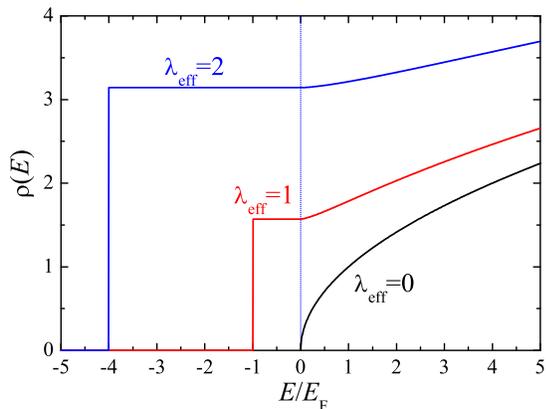} 
\par\end{centering}

\caption{(color online). Density of states of a 3D homogeneous SO coupled system
at several SO coupling strengths. The density of state is plotted
in units of $Mk_{F}/\hbar^{2}$.}

\label{figA1} 
\end{figure}

\end{document}